\documentclass[showpacs,preprintnumbers,amsmath,amssymb]{revtex4}

\usepackage{graphicx}
\usepackage{dcolumn}
\usepackage{bm}
\usepackage{natbib}
\usepackage{color}

\begin{document}


\title{The Weibull - log Weibull distribution for interoccurrence times of earthquakes}

\author{Tomohiro Hasumi}
 \email{t-hasumi.1981@toki.waseda.jp}
\author{Takuma Akimoto}
\author{Yoji Aizawa}
\affiliation{Department of Applied Physics, Advanced School of Science and Engineering, Waseda University, Tokyo 169-8555, Japan}

\date{\today}

\begin{abstract}
By analyzing the Japan Meteorological Agency (JMA) seismic catalog for different tectonic settings, we have found that the probability distributions of time intervals between successive earthquakes \textemdash interoccurrence times\textemdash  can be described by the superposition of the Weibull distribution and the log-Weibull distribution. 
In particular, the distribution of large earthquakes obeys the Weibull distribution with the exponent $\alpha_1 <1$, indicating the fact that the sequence of large earthquakes is not a Poisson process. 
It is found that the ratio of the Weibull distribution to the probability distribution of the interoccurrence time gradually increases with increase in the threshold of magnitude. 
Our results infer that Weibull statistics and log-Weibull statistics coexist in the interoccurrence time statistics, and that the change of the distribution is considered as the change of the dominant distribution. 
In this case, the dominant distribution changes from the log-Weibull distribution to the Weibull distribution, allowing us to reinforce the view that the interoccurrence time exhibits the transition from the Weibull regime to the log-Weibull regime. 
\end{abstract}

\maketitle

\section{\label{intro}Introduction}
Earthquakes are phenomena exhibiting great complexity characterized by many empirical statistical laws~\cite{Main:RG1996}. 
The time intervals between successive earthquakes can be classified into two types: interoccurrence times and recurrence times~\cite{Abaimov:NPG2007}. 
Interoccurrence times are the interval times between earthquakes on all faults in a region, whereas recurrence times are the time intervals between earthquakes in a single fault or fault segment. 
For a seismologist, recurrence times denote the interval times of characteristic earthquakes that occur quasi-periodically in a single fault. \par
Recently, a unified scaling law of interoccurrence times was reported using the Southern California~\cite{Bak:PRL2002} and worldwide earthquake catalogs \cite{Corral:PRL2004}, where the interoccurrence times were analyzed for the events with the magnitude $m$ above a certain threshold $m_c$ under the following two conditions: 
(a) earthquakes can be considered as a point process in space and time; 
(b) there is no distinction between foreshocks, mainshocks, and aftershocks. 
It has been demonstrated that the probability distribution of the interoccurrence time is well-fitted by the generalized gamma distribution. 
This scaling law is obtained by analyzing the aftershock data~\cite{Shcherbakov:PRL2005} and is derived approximately from a theoretical framework proposed by Saichev and Sornette~\cite{Saichev:PRL2006}. 
Abe and Suzuki showed that the survivor function of the interoccurrence time can be described by a power law~\cite{Abe:PA2005}. 
It has been reported that the sequence of aftershocks and successive independent earthquakes is a Poisson process~\cite{Gardner:BSSA1974, Enescu:GJI2008}. 
Recent works on interoccurrence time statistics are focused on the effect of \textquotedblleft long-term memory"~\cite{Bunde:PRL2005, Livina:PRL2005, Lennartz:EPL2008} as well as on the determination of the distribution function. 
However, the effect of changing the threshold of magnitude on the interoccurrence time statistics has not been discussed fully in the existing literature.  \par 
In this work, we trace the change in interoccurrence time statistics produced by varying of the cutoff magnitude $m_c$, which has not been studied previously. 
This study aims to infer the interoccurrence time statistics for middle or big mainshocks. 
It demonstrates that the interoccurrence time distribution is described by the superposition of the Weibull distribution and the log-Weibull distribution.
In particular, the distribution of large earthquakes follows the Weibull distribution with exponent $\alpha_1 <1$, indicating that the sequence of large earthquakes is not a Poisson process. 
In addition, the ratio of the Weibull distribution to the interoccurrence time distribution gradually increases as $m_c$ is increased. 
Our results lead to the conclusion that the interoccurrence time statistics follow Weibull statistics and log-Weibull statistics closely, and the change of the distribution can be interpreted as the change of a predominant distribution, i.e., the predominant distribution changes from the log-Weibull distribution to the Weibull distribution when $m_c$ is increased, so that the interoccurrence time statistics exhibit the Weibull - log Weibull transition. 

\section{Data Analysis and applicable distributions}
\begin{figure*}
\begin{center}
\includegraphics[width=.8\linewidth]{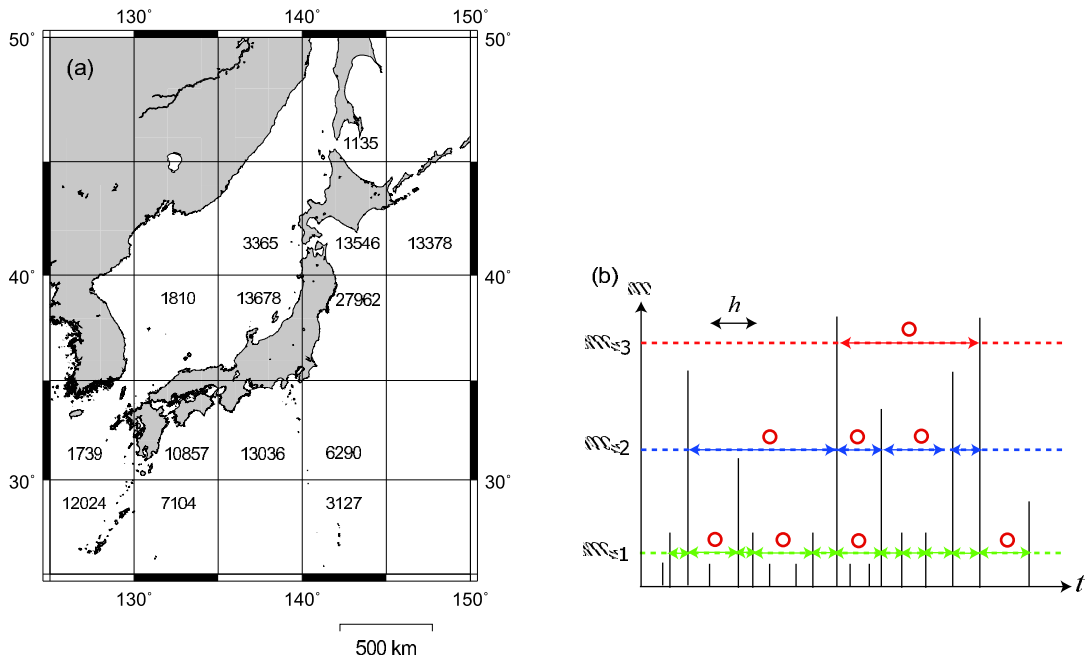}
\end{center}
\caption{(a): A map around Japan where we have carried out analysis the JMA catalog. The number of each bin stands for the number of earthquakes from 01/01/2001 to 10/31/2007. (b) Illustration of interoccurrence times for three different threshold of magnitude, $m_{c1}, m_{c2}$, and $m_{c3}$. In our work, we focus on interoccurrence times in the time domain, $\tau >h$, corresponding to $\circ$ in the figure.}
\label{time}
\end{figure*}

In this study, we investigate interoccurrence time statistics using the earthquake catalog made available by the Japan Meteorological Agency (JMA)~\cite{JMA}. 
This catalog contains occurrence times and hypocenter locations of earthquakes having a magnitude $m$ greater than 2.0, and covers the region spanning 25$^{\circ}$\textendash 50$^{\circ}$ N latitude and 125$^{\circ}$\textendash 150$^{\circ}$ E longitude (see Fig.~\ref{time} (a)). 
In this study, we use the data from January 1, 2001 to October 31, 2007. \par

Our method is similar to the that of previous works~\cite{Bak:PRL2002, Corral:PRL2004, Lennartz:EPL2008}~(see Fig.~\ref{time}); 
\begin{enumerate}
\item We divided the spatial areas into a window of $L$ degrees in longitude and $L$ degrees in latitude.
\item For each bin, earthquakes with magnitude $m$ above a certain cutoff magnitude $m_c$ were considered.
\item We analyzed the interoccurrence times and then performed the data fitting in the time domain, $\tau>h$ (day).  
\end{enumerate}
For each bin, we analyzed interoccurrence times using at least 100 events to avoid statistical errors.  
$h$ and $L$ are taken to be 0.5 and 5, respectively. 
As shown in Fig.~\ref{time} (a), we investigated the interoccurrence time statistics for different 14 regions.  
Aftershocks might be excluded from the study based on the information from previous studies~\cite{Corral:PRL2004, Enescu:GJI2008}. \par
One of our main goals in this study is to determine the distribution function of the interoccurrence time. 
Here, we will focus our attention on the applicability of the Weibull distribution $P_{w}$, the log-Weibull distribution $P_{lw}$~\cite{Huillet:EPJB1999}, the power law $P_{pow}$~\cite{Abe:PA2005}, the gamma distribution $P_{gam}$ (in the case of $\delta=1$ in the paper \cite{Corral:PRL2004}), and the log normal distribution $P_{ln}$~\cite{Matthews:BSSA2002}, which are defined as
\begin{eqnarray}
P_w(\tau) = {\displaystyle \left(\frac{\tau}{\beta_1} \right)^{\alpha_1 -1} \frac{\alpha_1}{\beta_1} \exp \left[-\left(\frac{\tau}{\beta_1}\right)^{\alpha_1} \right]}, \; \; 
P_{lw}(\tau) = {\displaystyle \frac{(\log (\tau /h))^{\alpha_2 -1}}{(\log \beta_2 )^{\alpha_2}} \frac{\alpha_2}{\tau} ,
\exp \left[-\left(\frac{\log (\tau /h)}{\log \beta_2}\right)^{\alpha_2}  \right]}\nonumber \\
P_{pow}(\tau) = {\displaystyle \frac{\beta_3 (\alpha_3 -1)}{(1+\beta_3 \tau)^{\alpha_3}}}, \; \; 
P_{gam}(\tau) = {\displaystyle \tau^{\alpha_4 -1} \frac{\exp{(-\tau/\beta_4)}}{\Gamma(\alpha_4) {\beta_4}^{\alpha_4}}}, \; \; 
P_{ln}(\tau) = {\displaystyle \frac{1}{\tau \beta_5\sqrt{2\pi}}\exp \left[-\frac{(\ln(\tau)-\alpha_5 )^2}{2\beta_5 ^2}\right]}
\end{eqnarray}
where $\alpha_{i}, \beta_{i}$, and $h$ are constants and characterize the distribution. 
$\Gamma(x)$ is the gamma function. 
$i$ stands for an index number; $i=1, 2, 3, 4$, and 5 correspond to the Weibull distribution, the log-Weibull distribution, the power law, the gamma distribution, and the log normal distribution, respectively. \par

The Weibull distribution is well known as a description of the probability distribution of failure-occurrence times. 
In seismology, the distribution of ultimate strain \cite{Hagiwara:TP1974}, the recurrence time distribution~\cite{Bakun:Nature2005, Abaimov:GJI2007}, and the damage mechanics of rocks \cite{Nanjo:JGR2005} show the Weibull distribution. 
In numerical studies, the recurrence time distribution in the 1D \cite{Abaimov:NPG2007} and 2D \cite{Hasumi:2008b} spring-block model, and  in the \textquotedblleft Virtual California model" \cite{Yakovlev:BSSA2006} also exhibit the Weibull distribution. 
It is known that for $\alpha_1 = 1$ and $\alpha_1 < 1$, the tail of the Weibull distribution is equivalent to the exponential distribution and  the stretched exponential distribution, respectively. 
The log-Weibull distribution is constructed by a logarithmic modification of the cumulative distribution of the Weibull distribution. 
In general, the tail of the log-Weibull distribution is much longer than that of the Weibull distribution. 
As for $\alpha_2 = 1$ the log-Weibull distribution is equal to the power law. 
It has been reported that the log-Weibull distribution can be derived from the chain-reaction model proposed by Huillet and Raynaud~\cite{Huillet:EPJB1999}. \par
In order to detect whether a specific distribution is preferred, we used three different goodness-of-fit tests: the root mean square (rms) test, the Kolmogorov-Smirnov (KS) test, and the Anderson-Darling (AD) test.
Firstly, we use the rms test. 
The rms value is defined as 
\begin{eqnarray}
{\rm rms} = \sqrt{\frac{\sum_{i=1} ^{n'} (x_i - x'_i)^2}{n'-k}},
\end{eqnarray}
where $x_i$ are actual data of the distribution and $x'_i$ are predicted data obtained from the ideal curve. 
$n$ and $k$ stand for the number of data point and fitting parameters, respectively. 
In this study, the rms value is calculated using the cumulative distribution function (cdf) and the Weibull plot. 
As readily known, the most appropriate distribution shows the smallest rms value.
Secondly, the KS test is performed. 
The Kolmogorov-Smirnov statistic $D_{KS}$ is defined by 
\begin{eqnarray}
D_{KS} = \max_i |y_i -y_i'|,
\end{eqnarray}
where $y_i$ and $y_i'$ mean the actual data of the cumulative distribution function and the data estimated from the fitted distribution, respectively. 
It is well recognized that the preferred distribution has the smallest value of $D_{KS}$.
Finally, the AD test is used. 
This test gives more weight to the tails of the distribution than the KS test. 
The Anderson-Darling statistic $A^2$ is defined as 
\begin{eqnarray}
A^2 = -N - \frac{1}{N} \sum _{i=1} ^{N} (2i -1) [\ln (F(z_i))+ \ln (1-F(z_{n-i+1}))],
\end{eqnarray}
where $N$ is the sample number, $F(x)$ is the cumulative distribution function, and $z_i$ is the $i$th interoccurrence time. Note that $z$ is put in order.
The most suitable distribution exhibits the smallest $A^2$.
We applied these goodness-of-fit tests by use of the five distributions which were mentioned before. 
\section{Results}
\begin{figure*}
\begin{center}
\includegraphics[width=.95\linewidth]{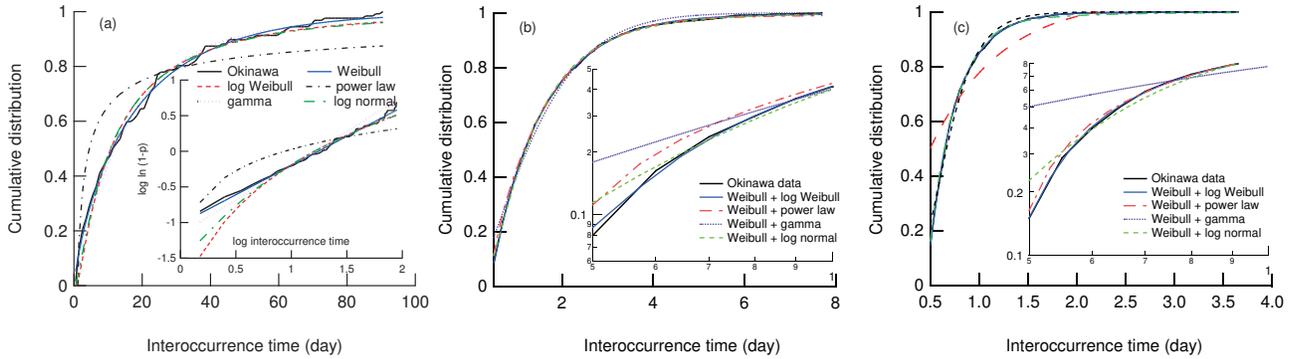}
\end{center}
\caption{Distribution of interoccurrence time at Okinawa region for different $m_c$ and distribution functions. (a) $m_c=4.5$, (b) $m_c=3.0$, and (c) $m_c=2.0$. Inset figure of (a) shows the Weibull plot and (b) and (c) represent the log-log scale of the figures.}
\label{nature}
\end{figure*}

\begin{table}
\caption{\label{table0}The results of the rms value, $D_{KS}$, $A^2$, and optimal parameters for different distribution functions in Fig.~\ref{nature} (a). The error bars are 95 per cent confidence limits.}
\begin{center}
\begin{tabular}{c|c|c|c|c|c|c|c}
 &distribution&$\alpha_i$ &$\beta_i$& rms (cdf) & $D_{KS}$ & $A^2$ & rms (Weibull-plot)  \\
\hline
\hline
  & $P_{w}~(i=1)$ & 0.82$\pm$ 0.01 & 17.5$\pm$ 0.28 & 0.014 & 0.039 & 0.89 & 0.065 \\
  $m_c=4.5$ & $P_{lw}~(i=2)$ & 2.95$\pm$ 0.12 & 32.4 $\pm$ 1.08 & 0.028 & 0.107 & 1.90 & 0.099  \\
 127 events & $P_{pow}~(i=3)$ & 1.46 $\pm$ 0.04 & 0.99$\pm$ 0.24 & 0.112 & 0.384 & 14.8 & 0.19 \\
  91 data points & $P_{gam}~(i=4)$ & 0.94$\pm$ 0.01 & 17.6 $\pm$ 0.44 & 0.022 & 0.066 & 1.51 & 0.098  \\
  & $P_{ln}~(i=5)$ & 2.37$\pm$ 0.04 & 1.22 $\pm$ 0.05 & 0.027 & 0.088 & 1.06  & 0.093  \\
\hline
\end{tabular}
\end{center}
\end{table}

\begin{figure*}
\begin{center}
\includegraphics[width=.95\linewidth]{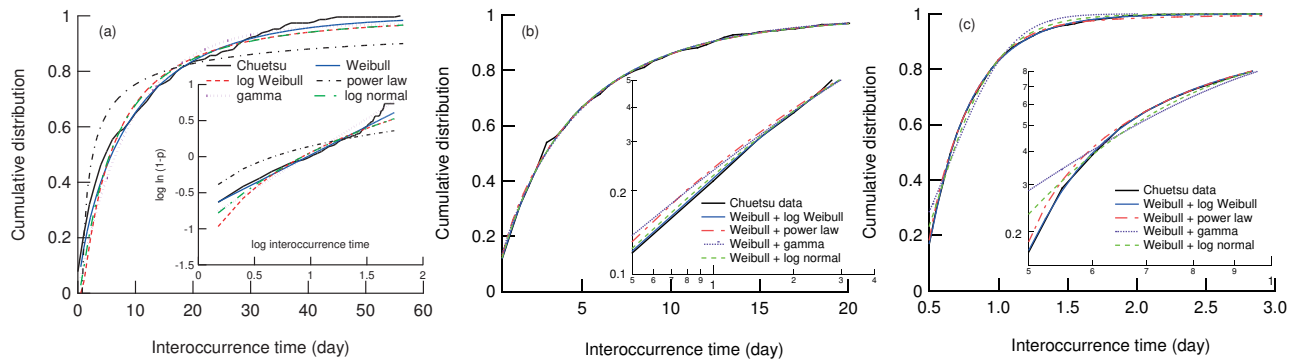}
\end{center}
\caption{Cumulative distribution of interoccurrence time for Chuetsu area at different $m_c$ and distribution function. (a), (b), and (c) represent interoccurrence time when $m_c=4.0$, $m_c=3.0$, and $m_c=2.0$, respectively. Inset figure of (a) shows the Weibull plot, and (b) and (c) represent log-log scale of the figures.}
\label{nigata}
\end{figure*}

\begin{table}[h]
\caption{\label{table0-1}The results of the rms value, $D_{KS}$, $A^2$, and optimal parameters for different distribution functions in Fig.~\ref{nigata} (a). The error bars are 95 per cent confidence limits.}
\begin{center}
\begin{tabular}{c|c|c|c|c|c|c|c}
&distribution&$\alpha_i$ &$\beta_i$& rms (cdf)  & $D_{KS}$ & $A^2$ & rms (Weibull-plot) \\
\hline
\hline
  & $P_{w}~(i=1)$ & 0.79$\pm$ 0.02 & 9.35 $\pm$ 0.28 & 0.019 & 0.034 & 3.17 & 0.083  \\
 $m_c=4.0$ & $P_{lw}~(i=2)$ & 2.36 $\pm$ 0.18 & 16.9 $\pm$ 1.08 & 0.038 & 0.106 & 3.71 & 0.134 \\
 231 events & $P_{pow}~(i=3)$ & 1.52 $\pm$ 0.06 & 0.68 $\pm$ 0.14 & 0.100 & 0.251 & 14.1 & 0.227\\
 57 data points & $P_{gam}~(i=4)$ & 0.95 $\pm$ 0.02 & 9.70 $\pm$ 0.48 & 0.034 & 0.072 & 3.41 & 0.077\\
  & $P_{ln}~(i=5)$ & 1.71 $\pm$ 0.06 & 1.27 $\pm$ 0.07 & 0.033 & 0.056 & 1.33 & 0.124 \\
\hline
\end{tabular}
\end{center}
\end{table}

\begin{table}
\caption{\label{table1}The interoccurrence time statistics of earthquakes in Okinawa region. The error bars are 95 per cent confidence limits.}
\begin{center}
\begin{tabular}{c|c|cc|cc|c|cc}
\multicolumn{1}{c}{$m_c$} & \multicolumn{1}{c}{Distribution X} &\multicolumn{2}{c}{Weibull distribution} & \multicolumn{2}{c}{Distribution X} & \multicolumn{1}{c}{Weibull rate} & \multicolumn{2}{c}{rms-value} \\
Region & index $i$ &$\alpha_1$ &$\beta_1$&$\alpha_i$&$\beta_i$ &$p$&  [$\times 10^{-3}$] & $\ln$ rms \\
\hline
\hline
 & $P_{lw}~(i=2)$ & 0.82 $\pm$ 0.01 & 17.5 $\pm$ 0.28 & $-$ & $-$ & 1 & 14 & $-4.27$ \\
4.5 & $P_{pow}~(i=3)$ & 0.82 $\pm$ 0.01 & 17.5 $\pm$ 0.28 & $-$ & $-$ & 1 & 14 & $-4.27$ \\
Okinawa & $P_{gam}~(i=4)$ & 0.82 $\pm$ 0.01 & 17.5 $\pm$ 0.28 & $-$ & $-$ & 1 & 14 & $-4.27$ \\
 & $P_{ln}~(i=5)$ &0.82 $\pm$ 0.01 & 17.5 $\pm$ 0.28 & $-$ & $-$ & 1 & 14 & $-4.27$ \\
\hline
 & $P_{lw}~(i=2)$ & 0.91 $\pm$ 0.01 & 8.28 $\pm$ 0.12 & $-$ & $-$ & 1 & 11 & $-4.51$ \\
4.0 & $P_{pow}~(i=3)$ & 0.91 $\pm$ 0.01 & 8.28 $\pm$ 0.12 & $-$ & $-$ & 1 & 11 & $-4.51$ \\
Okinawa & $P_{gam}~(i=4)$ & 0.91 $\pm$ 0.01 & 8.28 $\pm$ 0.12 & $-$ & $-$ & 1 & 11 & $-4.51$ \\
 & $P_{ln}~(i=5)$ & 0.91 $\pm$ 0.01 & 8.28 $\pm$ 0.12 & $-$ & $-$ & 1 & 11 & $-4.51$ \\
\hline
 & $P_{lw}~(i=2)$~\footnotemark[1] & 1.09 $\pm$ 0.03 & 3.74 $\pm$ 0.13  & 1.87 $\pm$ 0.25 & 4.73 $\pm$ 0.70 & 0.78 $\pm$ 0.03 & 5.1 & $-5.28$\\
3.5 & $P_{pow}~(i=3)$~\footnotemark[2] & 1.09 $\pm$ 0.03 & 3.74 $\pm$ 0.14 & 1.81 $\pm$ 0.07 & 0.64 $\pm$ 0.07 & 0.95 $\pm$ 0.02 & 7.3 & $-4.92$ \\
Okinawa & $P_{gam}~(i=4)$~\footnotemark[2] & 1.09 $\pm$ 0.03 & 3.74 $\pm$ 0.14  & $-$ & $-$ & 1 & 12 & $-4.42$ \\
 & $P_{ln}~(i=5)$~\footnotemark[2] & 1.09 $\pm$ 0.03 & 3.74 $\pm$ 0.14 & 0.86 $\pm$ 0.01 & 0.93 $\pm$ 0.01 & 0.66 $\pm$ 0.06 & 5.7 & $-5.17$ \\
\hline
 & $P_{lw}~(i=2)$~\footnotemark[1] & 1.43 $\pm$ 0.04 & 1.81 $\pm$ 0.04 & 1.34 $\pm$ 0.06 & 2.46 $\pm$ 0.08 & 0.60 $\pm$ 0.02 & 3.8 & $-5.57$ \\
3.0 & $P_{pow}~(i=3)$~\footnotemark[2] & 1.38 $\pm$ 0.04 & 1.64 $\pm$ 0.02 & 2.20 $\pm$ 0.08 & 0.55 $\pm$ 0.02 & 0.79 $\pm$ 0.02 & 8.8 & $-4.73$ \\
Okinawa & $P_{gam}~(i=4)$~\footnotemark[2] & 1.38 $\pm$ 0.04 & 1.64 $\pm$ 0.02  & $-$ & $-$ & 1 & 17 & $-4.07$ \\
 & $P_{ln}~(i=5)$~\footnotemark[2] &1.38 $\pm$ 0.04 & 1.64 $\pm$ 0.02 & 0.19 $\pm$ 0.006 & 1.46 $\pm$ 0.01 & 0.08 $\pm$ 0.08 & 6.4 & $-5.05$ \\
\hline
 & $P_{lw}~(i=2)$~\footnotemark[1] & 1.76 $\pm$ 0.04 & 1.16 $\pm$ 0.02 & 1.23 $\pm$ 0.03 & 1.67 $\pm$ 0.02 & 0.47 $\pm$ 0.01 & 3.0 & $-5.81$ \\
2.5 & $P_{pow}~(i=3)$~\footnotemark[2] & 1.90 $\pm$ 0.09 & 1.02 $\pm$ 0.01 & 2.83 $\pm$ 0.08 & 0.51 $\pm$ 0.01 & 0.59 $\pm$ 0.04 & 9.8 & $-4.63$ \\
Okinawa & $P_{gam}~(i=4)$~\footnotemark[2] & 1.90 $\pm$ 0.05 & 1.02 $\pm$ 0.008 & 1.09 $\pm$ 0.03 & 0.94 $\pm$ 0.04 & 0.99 $\pm$ 0.09 & 23 & $-3.77$ \\
 & $P_{ln}~(i=5)$ & $-$ & $-$ & $-$0.19 $\pm$ 0.008 & 0.53 $\pm$ 0.01 & 0 & 12 & $-4.42$ \\
\hline
 & $P_{lw}~(i=2)$~\footnotemark[1]  & 1.75 $\pm$ 0.06& 0.78 $\pm$ 0.02 & 1.18 $\pm$ 0.03 & 1.43 $\pm$ 0.02 & 0.42 $\pm$ 0.02 & 2.3 & $-6.07$ \\
2.0 & $P_{pow}~(i=3)$~\footnotemark[2] & 2.56 $\pm$ 0.18 & 0.77 $\pm$ 0.01 & 3.60 $\pm$ 0.09 & 0.48 $\pm$ 0.004 & 0.39 $\pm$ 0.04 & 7.3 & $-4.92$ \\
Okinawa & $P_{gam}~(i=4)$~\footnotemark[2]  & 2.56 $\pm$ 0.18 & 0.77 $\pm$ 0.01  & 1.09 $\pm$ 0.06 & 0.68 $\pm$ 0.05 & 0.96 $\pm$ 0.08 & 25 & $-3.09$ \\
 & $P_{ln}~(i=5)$ & $-$ & $-$ & $-$0.41 $\pm$ 0.01 & 0.39 $\pm$ 0.03 & 0 & 15 & $-4.20$ \\
\hline
\end{tabular}
\footnotetext[1]{We used the parameter estimation procedure (A).}
\footnotetext[2]{We used the parameter estimation procedure (B).}
\end{center}
\end{table}

\begin{table}
\caption{\label{table2}The interoccurrence time statistics of earthquakes in Chuetsu area. The error bars are 95 per cent confidence limits.}
\begin{center}
\begin{tabular}{c|c|cc|cc|c|cc}
\multicolumn{1}{c}{$m_c$} & \multicolumn{1}{c}{Distribution X} &\multicolumn{2}{c}{Weibull distribution} & \multicolumn{2}{c}{Distribution X} & \multicolumn{1}{c}{Weibull rate} & \multicolumn{2}{c}{rms-value} \\
Region & index $i$ &$\alpha_1$ &$\beta_1$&$\alpha_i$&$\beta_i$ &$p$&  [$\times 10^{-3}$] & $\ln$ rms \\
\hline
\hline
 & $P_{lw}~(i=2)$ & 0.79 $\pm$ 0.02 & 9.35 $\pm$ 0.28 & $-$ & $-$ & 1 & 19 & $-3.96$ \\
4.0 & $P_{pow}~(i=3)$ & 0.79 $\pm$ 0.02 & 9.35 $\pm$ 0.28 & $-$ & $-$ & 1 & 19 & $-3.96$ \\
Chuetsu & $P_{gam}~(i=4)$ &  0.79 $\pm$ 0.02 & 9.35 $\pm$ 0.28 & $-$ & $-$ & 1 & 19 & $-3.96$\\
 & $P_{ln}~(i=5)$ &  0.79 $\pm$ 0.02 & 9.35 $\pm$ 0.28 & $-$ & $-$ & 1 & 19 & $-3.96$ \\
\hline
 & $P_{lw}~(i=2)$~\footnotemark[2] & 0.85 $\pm$ 0.007 & 4.56 $\pm$ 0.03 & 1.96 $\pm$ 0.03 & 8.16 $\pm$ 0.19 & 0.86 $\pm$ 0.02 & 6.2 &$-5.08$ \\
3.5 & $P_{pow}~(i=3)$~\footnotemark[2] & 0.85 $\pm$ 0.007 & 4.56 $\pm$ 0.03 & 1.66 $\pm$ 0.03 & 0.60 $\pm$ 0.05 & 0.96 $\pm$ 0.01 & 7.1 & $-4.95$\\
Chuetsu & $P_{gam}~(i=4)$~\footnotemark[2] & 0.85 $\pm$ 0.007 & 4.56 $\pm$ 0.03 & 0.96 $\pm$ 0.004 & 4.61 $\pm$ 0.05 & 0.92 $\pm$ 0.07 & 7.6 & $-4.88$ \\
 & $P_{ln}~(i=5)$~\footnotemark[2] & 0.85 $\pm$ 0.007 & 4.56 $\pm$ 0.03 & 1.05 $\pm$ 0.02 & 2.30 $\pm$ 0.04 & 0.78 $\pm$ 0.04 & 6.4 & $-5.05$ \\
\hline
 & $P_{lw}~(i=2)$~\footnotemark[2] & 1.08 $\pm$ 0.02 & 2.17 $\pm$ 0.02  & 1.99 $\pm$ 0.16 & 5.35 $\pm$ 0.39 & 0.82 $\pm$ 0.04 & 3.9 & $-5.55$ \\
3.0 & $P_{pow}~(i=3)$~\footnotemark[2] & 1.08 $\pm$ 0.02 & 2.17 $\pm$ 0.02 & 1.97 $\pm$ 0.04 & 0.53 $\pm$ 0.02 & 0.93 $\pm$ 0.009 & 5.0 & $-5.30$ \\
Chuetsu & $P_{gam}~(i=4)$~\footnotemark[2] & 1.08 $\pm$ 0.02 & 2.17 $\pm$ 0.02  & $-$ & $-$ & 1 & 6.5 & $-5.04$ \\
 & $P_{ln}~(i=5)$~\footnotemark[2] & 1.08 $\pm$ 0.02 & 2.17 $\pm$ 0.02  & 0.40 $\pm$ 0.02 & 0.92 $\pm$ 0.04 & 0.65 $\pm$ 0.06 & 3.7 & $-5.60$ \\
\hline
 & $P_{lw}~(i=2)$~\footnotemark[1] & 1.47 $\pm$ 0.03 & 1.24 $\pm$ 0.02 & 1.20 $\pm$ 0.04 & 1.91 $\pm$ 0.04 & 0.59 $\pm$ 0.01 & 2.4 & $-6.03$ \\
2.5 & $P_{pow}~(i=3)$~\footnotemark[2] & 1.55 $\pm$ 0.06 & 1.17 $\pm$ 0.02 & 2.52 $\pm$ 0.08 & 0.50 $\pm$ 0.02 & 0.69 $\pm$ 0.03 & 6.8 & $-4.99$ \\
Chuetsu & $P_{gam}~(i=4)$~\footnotemark[2] & 1.55 $\pm$ 0.06 & 1.17 $\pm$ 0.02 & 1.03 $\pm$ 0.02 & 1.10 $\pm$ 0.05 & 0.99 $\pm$ 0.09 & 16 & $-4.14$ \\
 & $P_{ln}~(i=5)$ & $-$ & $-$ & $-$0.10 $\pm$ 0.006 & 0.64 $\pm$ 0.008 & 0 & 5.1 & $-5.28$ \\
\hline
 & $P_{lw}~(i=2)$~\footnotemark[1] & 1.77 $\pm$ 0.08 & 0.78 $\pm$ 0.02 & 1.20 $\pm$ 0.04 & 1.47 $\pm$ 0.02 & 0.47 $\pm$ 0.02 & 2.8 & $-5.88$ \\
2.0 & $P_{pow}~(i=3)$~\footnotemark[2] & 2.43 $\pm$ 0.18 & 0.79 $\pm$ 0.01 & 3.47 $\pm$ 0.12 & 0.48 $\pm$ 0.008 & 0.47 $\pm$ 0.05 & 8.0 & $-4.83$ \\
Chuetsu & $P_{gam}~(i=4)$~\footnotemark[2] & 2.43 $\pm$ 0.18 & 0.79 $\pm$ 0.01  & 1.19 $\pm$ 0.11 & 0.69 $\pm$ 0.04 & 0.97 $\pm$ 0.11 & 26 & $-3.65$ \\
 & $P_{ln}~(i=5)$ & $-$ & $-$ & $-$0.40 $\pm$ 0.005 & 0.82 $\pm$ 0.02& 0 & 14 & $-4.26$ \\
\hline
\end{tabular}
\footnotetext[1]{We used the parameter estimation procedure (A).}
\footnotetext[2]{We used the parameter estimation procedure (B).}
\end{center}
\end{table}

The cumulative distributions of the interoccurrence time for different $m_c$ in Okinawa region (125$^{\circ}$\textendash 130$^{\circ}$E and 25$^{\circ}$\textendash 30$^{\circ}$N) and in Chuetsu region (135$^{\circ}$\textendash 140$^{\circ}$E and 35$^{\circ}$\textendash 40$^{\circ}$N) are displayed in Figure~\ref{nature} and \ref{nigata}, respectively. 
The total number of earthquakes were 12024 in Okinawa and 13678 in Chuetsu. \par 
At first, we focus on the interoccurrence time distribution for large $m_c$. 
We have tried four statistical tests, the rms (cdf), the KS test, the AD test, and the rms (Weibull-plot) test, whose results are shown in Table~\ref{table0} for Okinawa and Table~\ref{table0-1} for Chuetsu. 
For Okinawa, we can certify that the most suitable distribution is the Weibull distribution in all tests. 
In general, there is a possibility that the preferred distribution is not unique but depends on the test we use. 
However, the results obtained in Table~\ref{table0} seems to support that the Weibull distribution is the most appropriate distribution in this case.  
As for Chuetsu, by four tests, the preferred distribution is suited to be the Weibull distribution as shown in Table~\ref{table0-1}, where the Weibull distribution is the most prominent distribution in the two tests (rms (cdf) and KS test), although the Weibull distribution is not the most appropriate distribution in the remaining two tests (AD test and rms (Weibull-plot)).  
Thus, we reinforce the view that the Weibull distribution is preferred. 
Hereinafter the preferred distribution function is evaluated by use of the rms (cdf) test. \par

However, the fitting accuracy of the Weibull distribution becomes worse with a gradual decrease in $m_c$. 
We now propose a possible explanation that states that \textquotedblleft the interoccurrence time distribution can be described by the superposition of the Weibull distribution and another distribution, hereafter referred to as the distribution X, $P_{X}$",
\begin{equation}
P(\tau) = p \times P_{w} + (1-p) \times P_{X}, 
\label{eq.1}
\end{equation}
where $p$ is a parameter in the range, $0 \le p \le 1$ and stands for the ratio of $P_{w}$ divided by $P(\tau)$.  
The interoccurrence time distribution obeys the Weibull distribution for $p=1$, whereas it follows the distribution X for $p=0$. 
In this study, the log-Weibull distribution, the power law, the gamma distribution, and the log normal distribution are candidates for the distribution X. \par
We shall explain the parameter estimation procedure; (A) the optimal parameters are estimated so as to minimize the differences between the data and the test function by varying five parameters, $\alpha_1, \beta_1, \alpha_i, \beta_i$, and $p$. 
If there is a parameter, where $C_v$, the ratio of the standard deviation divided by the mean exceeds 0.1, another estimation procedure is performed. 
(B) the Weibull parameters, $\alpha_1$ and $\beta_1$, and the distribution X parameters, $\alpha_i$ and $\beta_i$, are optimized dependently and then $p$ is estimated. 


On the basis of this hypothesis, the fitting results of $P(\tau)$ are listed in Table~\ref{table1} for Okinawa region and in Table~\ref{table2} for Chuetsu region. 
We can assume that the Weibull distribution is the fundamental distribution, because $p=1$ for large $m_c$, which means that the effect of the distribution X is negligible. 
As observed in Table~\ref{table1} and \ref{table2}, the log-Weibull distribution is the most suitable distribution for the distribution X. 
Thus, we find that the interoccurrence times distribution can be described by the superposition of the Weibull distribution and the log-Weibull distribution, namely,
\begin{equation}
P(\tau) = p \times P_{w} + (1-p) \times P_{lw},
\end{equation}
$P(\tau)$ is controlled by five parameters, $\alpha_1, \alpha_2, \beta_1, \beta_2$, and $p$. \par

\begin{figure*}
\begin{center}
\includegraphics[width=.4\linewidth]{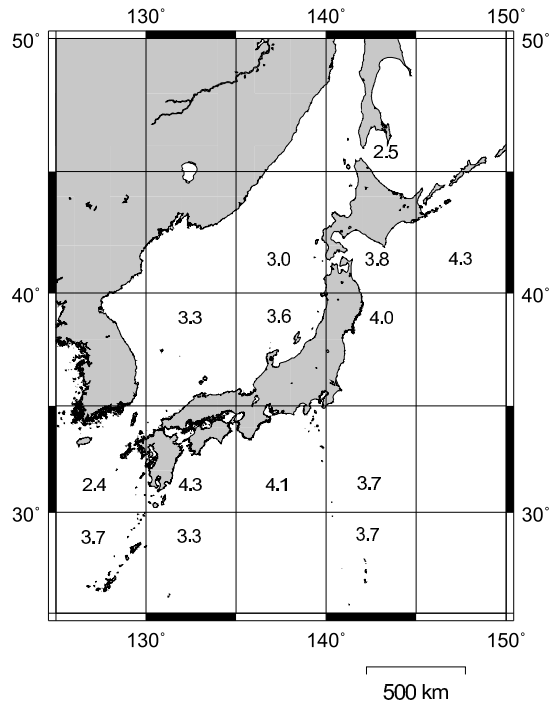}
\end{center}
\caption{The crossover magnitude $m_c^{**}$ map around Japan. $m_c^{**}$ depends on the region and ranges from 2.4 to 4.3.}
\label{crossover}
\end{figure*}

The result of fitting parameters of $P(\tau)$ for different regions are listed in Table~\ref{table3}. 
It was found that the interoccurrence time distribution of earthquakes with large $m_c$ obeys the Weibull distribution with the exponent $\alpha_1 < 1$, which has been observed previously. 
As shown in Tables~\ref{table1}, \ref{table2}, and \ref{table3}, we stress the point that the distribution function of the interoccurrence time changes with varying $m_c$.  
This indicates that the interoccurrence time statistics basically contain both Weibull and log-Weibull statistics, and the change of distribution function can be interpreted as the change of a dominant distribution. 
In this case, the dominant distribution of the interoccurrence time changes from the log-Weibull distribution to the Weibull distribution with $m_c$ increased. 
Thus, the point that the interoccurrence time statistics exhibit transition from the Weibull regime to the log-Weibull regime is reinforced. 
It is noted that a crossover magnitude from the superposition regime to the Weibull regime, denoted by $m_c^{**}$, depends on the spatial area on which we have focused in this study. 
We display $m_c^{**}$ map around Japan in Fig.~\ref{crossover}. 
$m_c^{**}$ ranges from 2.4 (125$^\circ$E\textendash 130$^\circ$E and 30$^\circ$N \textendash 35$^\circ$N) to 4.3 (130$^\circ$E\textendash 135$^\circ$E and 30$^\circ$N\textendash 35$^\circ$N, and 145$^\circ$E\textendash 150$^\circ$E and 40$^\circ$N\textendash 45$^\circ$N). 
Comparing Fig.~\ref{time} (a) with Fig.~\ref{crossover}, we found that a Weibull - log Weibull transition occurs in all region where we conducted. 

\begin{table}
\caption{\label{table3} Summary of the interoccurrence time statistics for different regions in Japan. The error bars are 95 per cent confidence limits.}
\begin{center}
\begin{tabular}{c|c|cc|cc|c}
\multicolumn{1}{c}{ } & \multicolumn{1}{c}{ } &\multicolumn{2}{c}{Weibull distribution} & \multicolumn{2}{c}{log-Weibull distribution} & \multicolumn{1}{c}{Weibull rate } \\
Region&$m_c$&$\alpha_1$&$\beta_1$ [day] &$\alpha_2$&$\beta_2$ [day]&$p$\\
\hline
\hline
 & 2.0~\footnotemark[1] & 1.26 $\pm$ 0.02 & 1.59 $\pm$ 0.05 & 1.43 $\pm$ 0.08 & 2.07 $\pm$ 0.05 & 0.59 $\pm$ 0.03 \\
 (A) & 2.5~\footnotemark[1] & 1.31 $\pm$ 0.02 & 1.90 $\pm$ 0.02 & 1.45 $\pm$ 0.09 & 2.09 $\pm$ 0.04 & 0.67 $\pm$ 0.02 \\
 140$^\circ$E\textendash 145$^\circ$E & 3.0~\footnotemark[2] & 1.13 $\pm$ 0.02 & 1.80 $\pm$ 0.02 & 1.51 $\pm$ 0.16 & 3.20 $\pm$ 0.10& 0.92 $\pm$ 0.04\\
 25$^\circ$N\textendash 30$^\circ$N& 3.5~\footnotemark[2] & 1.11 $\pm$ 0.01 & 2.49 $\pm$ 0.01 & 1.80 $\pm$ 0.13 & 1.47 $\pm$ 0.01& 0.95 $\pm$ 0.02 \\
 & 4.0 & 0.97 $\pm$ 0.01 & 4.45 $\pm$ 0.02 & \textendash & \textendash & 1 \\
 & 4.5 & 0.89 $\pm$ 0.02 & 11.3 $\pm$ 0.19 & \textendash & \textendash & 1 \\
\hline
 & 2.0~\footnotemark[2] & 4.34 $\pm$ 0.73 & 0.58 $\pm$ 0.02 & 1.04 $\pm$ 0.08 & 1.21 $\pm$ 0.01 & 0.40 $\pm$ 0.06 \\
 & 2.5~\footnotemark[1] & 2.91 $\pm$ 0.28 & 0.79 $\pm$ 0.02 & 1.10 $\pm$ 0.05 & 1.43 $\pm$ 0.02 & 0.31 $\pm$ 0.04 \\
  (B)  & 3.0~\footnotemark[1] & 1.54 $\pm$ 0.04 & 1.14 $\pm$ 0.04 & 1.37 $\pm$ 0.06 & 1.76 $\pm$ 0.06 & 0.62 $\pm$ 0.02 \\
 140$^\circ$E\textendash 145$^\circ$E & 3.5~\footnotemark[2] & 1.22 $\pm$ 0.02 & 1.82 $\pm$ 0.02 & 1.59 $\pm$ 0.14 & 3.27 $\pm$ 0.02 & 0.85 $\pm$ 0.04 \\
 35$^\circ$N\textendash 40$^\circ$N & 4.0 & 0.94 $\pm$ 0.009 & 3.45 $\pm$ 0.02 & \textendash & \textendash & 1 \\
 & 4.5 & 0.82 $\pm$ 0.02 & 7.69 $\pm$ 0.14 & \textendash & \textendash & 1 \\
 & 5.0 & 0.84 $\pm$ 0.03 & 18.7 $\pm$ 0.55 & \textendash & \textendash & 1 \\
\hline
 & 2.0~\footnotemark[1] & 2.17 $\pm$ 0.04 & 0.76 $\pm$ 0.006 & 1.11 $\pm$ 0.02 & 1.39 $\pm$ 0.02 & 0.52 $\pm$ 0.02 \\
 (C) & 2.5~\footnotemark[1] & 1.82 $\pm$ 0.05 & 1.04 $\pm$ 0.02 & 1.22 $\pm$ 0.04 & 1.66 $\pm$ 0.02 & 0.50 $\pm$ 0.02 \\
 135$^\circ$E\textendash 140$^\circ$E  & 3.0~\footnotemark[1] & 1.37 $\pm$ 0.03 & 1.71 $\pm$ 0.03 & 1.33 $\pm$ 0.05 & 2.24 $\pm$ 0.07 & 0.60 $\pm$ 0.02 \\
 30$^\circ$N\textendash 35$^\circ$N & 3.5~\footnotemark[2] & 1.01 $\pm$ 0.01 & 3.63 $\pm$ 0.06 & 1.90 $\pm$ 0.18 & 4.91 $\pm$ 0.55 & 0.87 $\pm$ 0.02 \\
 & 4.0~\footnotemark[2] & 0.88 $\pm$ 0.01 & 8.24 $\pm$ 0.09 & 2.52 $\pm$ 0.10 & 15.1 $\pm$ 0.47 & 0.95 $\pm$ 0.08 \\
 & 4.5 & 0.93 $\pm$ 0.01 & 21.6 $\pm$ 0.19 & \textendash & \textendash & 1 \\
\hline
\end{tabular}
\footnotetext[1]{We used the parameter estimation procedure (A).}
\footnotetext[2]{We used the parameter estimation procedure (B).}
\end{center}
\end{table}

\section{Discussion}

In our study of the size-dependence of the interoccurrence time statistics, the window size $L$ varied from 3$^{\circ}$ to 25$^{\circ}$. 
We used the data covering the region 140$^{\circ}$\textendash 143$^{\circ}$ E and 35$^{\circ}$\textendash 38$^{\circ}$ N for $L=3$, 140$^{\circ}$\textendash 145$^{\circ}$ E and 35$^{\circ}$\textendash 40$^{\circ}$ N for $L=5$, 140$^{\circ}$\textendash 150$^{\circ}$ E and 35$^{\circ}$\textendash 45$^{\circ}$ N for $L=10$, and 125$^{\circ}$\textendash 150$^{\circ}$ E and 25$^{\circ}$\textendash 50$^{\circ}$ N for $L=25$. 
For $L=25$, the data covers the entire region of the JMA catalog. 
The result of fitting parameters of $P(\tau)$, the crossover magnitude $m_c^{**}$, and the rms value are listed in Table~\ref{table4}. 
We have demonstrated that the Weibull exponent $\alpha_1$ is less than unity and the Weibull - log Weibull transition occurs in all cases. 
$m_c^{**}$ depends on $L$, namely $m_c^{**}=3.9$ for $L=3$, $m_c^{**}=4.0$ for $L=5$, $m_c^{**}=4.2$ for $L=10$, and $m_c^{**}=5.0$ for $L=25$.  
Therefore we can conclude that the interoccurrence time statistics presented are valid from $L=3$ to $L=25$. 


\begin{table}
\caption{\label{table4} The interoccurrence time statistics for different system size $L$. The error bars are 95 per cent confidence limits.}
\begin{center}
\begin{tabular}{c|c|c|c|c|c|c|c}
$L$ & Region & $m_c^{**}$ & $m_c$  & $\alpha_1$& $\beta_1$ [day]  & rms  & $\ln$ rms \\
\hline
$L=3$ & 140$^{\circ}$\textendash 143$^{\circ}$ E and 35$^{\circ}$\textendash 38$^{\circ}$ N & 3.9 & 4.6 & 0.88 $\pm$ 0.02 & 19.4 $\pm$ 0.36 & 0.011 & $-4.51$ \\
\hline
$L=5$ & 140$^{\circ}$\textendash 145$^{\circ}$ E and 35$^{\circ}$\textendash 40$^{\circ}$ N & 4.0 & 4.7 & 0.75 $\pm$ 0.03 & 10 $\pm$ 0.38 & 0.014 & $-4.27$  \\
\hline
$L=10$ & 140$^{\circ}$\textendash 150$^{\circ}$ E and 35$^{\circ}$\textendash 45$^{\circ}$ N & 4.2 & 4.9 & 0.94 $\pm$ 0.01 & 8.36 $\pm$ 0.08 & 0.0077  & $-4.87$  \\
\hline
$L=25$ & 125$^{\circ}$\textendash 150$^{\circ}$ E and 25$^{\circ}$\textendash 50$^{\circ}$ N & 5.0 & 5.7 & 0.93 $\pm$ 0.03 & 17.8 $\pm$ 0.40 & 0.021 & $-3.86$  \\
\hline
\end{tabular}
\end{center}
\end{table}

Finally, we compared our results with those of the previous studies. 
The unified scaling law shows a generalized gamma distribution which is approximately the gamma distribution, because $\delta$ in Corral's paper~\cite{Corral:PRL2004} is close to unity ($\delta =0.98 \pm 0.05$). 
For a long time domain, this distribution decays exponentially, supporting the view that an earthquake is a Poisson process. 
However, we have demonstrated that the Weibull distribution is more appropriate than the gamma distribution because the rms value obtained from the Weibull distribution is smaller than that from the gamma distribution. 
In addition, for large $m_c$, the probability distribution in a long time domain is similar to the stretched exponential distribution because $\alpha_1$ is less than unity, suggesting that earthquakes obey the long-tail distribution. 
We stress the point that the probability distribution changes by varying $m_c$, supporting the view that a transition occurs from the Weibull regime to the log-Weibull regime, which has not been reported previously.  
Recently, Abaimov {\it et al.} showed that the recurrence time distribution is also well-fitted by the Weibull distribution~\cite{Abaimov:NPG2007} rather than the Brownian passage time (BPT) distribution~\cite{Matthews:BSSA2002} and the log normal distribution. 
Taken together, we infer that both the recurrence time statistics and the interoccurrence time statistics show the Weibull distribution. \par

\begin{figure*}[]
\begin{center}
\includegraphics[width=.95\linewidth]{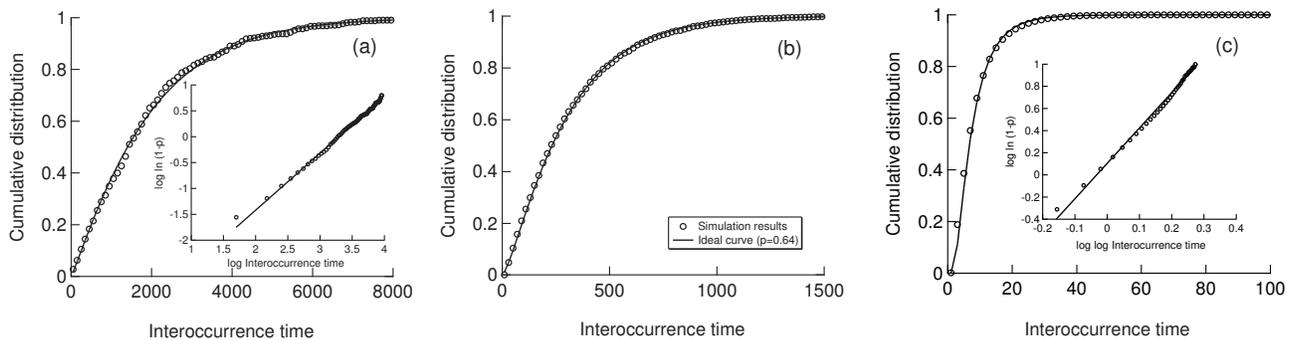}
\end{center}
\caption{The interoccurrence time distribution for different $m_c$ obtained from the spring-block model. More details are studied in Ref.~\cite{Hasumi:2008b}. The Weibull regime, the superposition regime, and the log-Weibull regime are shown in (a), (b), and (c), respectively. Inset figure of (a) and (c) are the Weibull-plot and log-Weibull-plot, respectively.}
\label{model}
\end{figure*}

In this study, we propose a new insight into the interoccurrence time statistics, stating that the interoccurrence statistics exhibit the Weibull - log Weibull transition.
This stresses that the distribution function can be described by the superposition of the Weibull distribution and the log-Weibull distribution, and that the predominant distribution function changes from the log-Weibull distribution to the Weibull distribution as $m_c$ is increased. 
Note that there is a possibility that a more suitable distribution might be found instead of the log-Weibull distribution. 
However, since our results are also obtained by analyzing the Southern California and the Taiwan earthquake catalogs as well \cite{Hasumi:2008a} we believe that the log-Weibull distribution is the best. 
Furthermore, the Weibull - log Weibull transition can be extracted more clearly by analyzing synthetic catalogs produced by the spring-block model (Fig.~\ref{model})~\cite{Hasumi:2008b}.  

\section{Conclusion}
In conclusion, we have proposed a new feature of interoccurrence time statistics by analyzing the JMA earthquake catalogs for different tectonic conditions. 
We found that the probability distribution of the interoccurrence time can be described clearly by the superposition of the Weibull distribution and the log-Weibull distribution. 
Especially for large earthquakes, the interoccurrence time distribution obeys the Weibull distribution with the exponent $\alpha_1 <1$, indicating that a large earthquake is not a Poisson process but a phenomenon exhibiting a long-tail distribution. 
As the threshold of magnitude $m_c$ increases, the ratio of the Weibull distribution to the interoccurrence time distribution $p$ gradually increases. 
Our findings support the view that the Weibull statistics and log-Weibull statistics coexist in the interoccurrence time statistics. 
We interpret the change of distribution function as the change of the predominant distribution function; the predominant distribution changes from the log-Weibull distribution and the Weibull distribution when $m_c$ is increased. 
Therefore, it is concluded that the interoccurrence time statistics exhibit the Weibull - log Weibull transition. 
We believe that this work is a first step toward a theoretical and geophysical understanding of this transition. 

\begin{acknowledgments}
We thank the JMA for allowing us to use the earthquake data. 
This work is partly supported by the Sasagawa Scientific Research Grant from The Japan Science Society. 
TH is grateful for research support from the Japan Society for the Promotion of Science (JSPS) and the Earthquake Research Institute cooperative research program at the University of Tokyo. 
Thanks are also extended to two anonymous reviewers for improving the manuscript. 
\end{acknowledgments}


\begin{thebibliography}{00}
\bibitem{Main:RG1996}
I. G. Main, Rev. Geophys., \textbf{34}, 433 (1996).
\bibitem{Abaimov:NPG2007}
S. G. Abaimov, D. L. Turcotte, R. Shcherbakov, and J. B. Rundle, Nonlinear Processes Geophys. {\bf 14}, 455 (2007).
\bibitem{Bak:PRL2002}
P. Bak, K. Christensen, L. Danon, and T. Scanlon, Phys. Rev. Lett., {\bf 88}, 178501 (2002). 
\bibitem{Corral:PRL2004}
A. Corral, Phys. Rev. Lett., {\bf 92}, 108501 (2004).
\bibitem{Shcherbakov:PRL2005}
R. Shcherbakov, G. Yakovlev, D. L. Turcotte, and J. B. Rundle, Phys. Rev. Lett., {\bf 95}, 218501 (2005).
\bibitem{Saichev:PRL2006}
A. Saichev and D. Sornette, Phys. Rev. Lett., {\bf 97}, 078501 (2006).
\bibitem{Abe:PA2005}
S. Abe and N. Suzuki, Physica A, {\bf 350}, 588, (2005).
\bibitem{Gardner:BSSA1974}
J. K. Gardner, L. Knopoff, Bull. Seismol. Soc. Am., {\bf 64}, 1363 (1974).
\bibitem{Enescu:GJI2008}
B. Enescu, Z. Struzik, and K. Kiyono, Geophys. J. Int, {\bf 172}, 395, (2008).
\bibitem{Bunde:PRL2005}
A. Bunde, J. F. Eichner, J. W. Kantelhardt, and S. Havlin, Phys. Rev. Lett. {\bf 94}, 048701 (2005).
\bibitem{Livina:PRL2005}
V. N. Livina, S. Havlin, and A. Bunde, Phys. Rev. Lett. {\bf 95}, 208501 (2005).
\bibitem{Lennartz:EPL2008}
S. Lennartz, V. N. Livina, A. Bunde, and S. Havlin, Europhys. Lett., {\bf 81}, 69001, (2008).
\bibitem{JMA}
Japan Meteorological Agency Earthquake Catalog: http://wwweic.eri.u-tokyo.ac.jp/db/jma1.
\bibitem{Hagiwara:TP1974}
Y. Hagiwara, Tectonophys, {\bf 23}, 313 (1974).
\bibitem{Bakun:Nature2005}
W. H. Bakun, B. Aagard, B. Dost, et al. Nature, {\bf 437}, 969, (2005).
\bibitem{Abaimov:GJI2007}
S. G. Abaimov, D. L. Turcotte, and J. B. Rundle, Geophys. J. Int. {\bf 170}, 1289 (2007).
\bibitem{Nanjo:JGR2005}
K. Z. Nanjo, D. L. Turcotte, and R. Shcherbakov, J. Geophys. Res., {\bf 110} B07403 (2005).
\bibitem{Hasumi:2008b}
T. Hasumi, T. Akimoto, and Y. Aizawa, Physica A, {\bf 388}, 483, (2009).
\bibitem{Yakovlev:BSSA2006}
G. Yakovlev, D. L. Turcotte, J. B. Rundle, and P. B. Rundle, Bull. Seismol. Soc. Am., {\bf 96}, 1995 (2006).
\bibitem{Hasumi:2008a}
T. Hasumi, C. Chen, T. Akimoto, and Y. Aizawa, arXiv:0808.2793.  
\bibitem{Huillet:EPJB1999}
T. Huillet and H. F. Raynaud, Eur. Phys. J. B., {\bf 12}, 457, (1999).  
\bibitem{Matthews:BSSA2002}
M. V. Matthews, W. L. Ellsworth, and A. P. Reasenberg, Bull. Seismol. Soc. Am., {\bf 92}, 2233, (2002). 

\end{thebibliography}
\end{document}